\documentclass[final,3p,times,twocolumn]{elsarticle}
\usepackage{graphicx}
\usepackage{nicefrac}
\usepackage{amsmath}
\usepackage{amssymb}
\usepackage{textcomp}
\usepackage{color}
\newcommand{\red}[1]{\textcolor{black}{#1}}
\usepackage{epstopdf}
\biboptions{comma,sort&compress,numbers}
\journal{Carbon}
\begin{document}

\begin{frontmatter}

\title{Magnetic order and superconductivity observed in
bundles of Double-Wall Carbon Nanotubes}
\author[exp]{J. Barzola-Quiquia}

\author[exp]{P. Esquinazi\corref{corr1}}
\cortext[corr1]{Corresponding author. Tel/Fax: +49 341 9732751/69.
E-mail address: esquin@physik.uni-leipzig.de (P. Esquinazi)}

\author[exp]{M. Lindel}
\address[exp]{Division of Superconductivity and
Magnetism, Institute for Experimental Physics II, University of
Leipzig, D-04103 Leipzig, Germany}  
\author[exp3]{D. Spemann}
\address[exp3]{Division Nuclear Solid State Physics, Institute for Experimental Physics II, University of
Leipzig, D-04103 Leipzig, Germany}
\author[exp2]{M. Muallem}
\author[exp2]{G.
D. Nessim} \address[exp2]{Department of Chemistry, Bar Ilan
Institute for Nanotechnology and Advanced Materials (BINA), Bar
Ilan University, 52900 Ramat Gan, Israel }

\begin{abstract}
The magnetotransport properties were studied in hundreds of
micrometer length double-wall carbon nanotubes (DWCNT) bundles.
Above 15~K the resistance shows an ohmic behavior and its
temperature dependence is well described using the variable-range
hopping for one-dimensional system. The magnetoresistance is
negative and can be explained using an empirical model based on
spin-scattering processes indicating the existence of magnetic
order up to room temperature. At temperatures between 2~K and 15~K
the resistance is non-ohmic and the current-voltage
characteristics reveal the appearance of a potential, which can be
well described by a fluctuation-induced tunneling conduction
model. In this low temperature range and at low enough input
current, a positive magnetoresistance appears - in addition to the
negative one - with an extraordinary hysteresis in field and
vanishes at $T \sim 15~$K, suggesting the existence of a
superconducting state. Magnetization results partially support the
existence of both phenomena in the DWCNT bundles.
\end{abstract}
\end{frontmatter}
\section{Introduction}
The search for  superconductivity and magnetic order  in
carbon-based materials triggered a large number of studies in
recent years. Experimental as well as theoretical works indicate
that magnetic order at high temperatures  in graphite is possible
through the influence of vacancies and/or hydrogen (for recent
reviews see \cite{yaz10,dim13} and refs. therein).  In contrast to
graphite and in spite of theoretical predictions on the
possibility to have magnetic order due to hydrogen or vacancies in
carbon nanotubes (CNT) \cite{ma04,ma05,sai05}, the observation of
this phenomenon in these carbon structures appears to be more
difficult. Apparently, only the hydrogenated CNT prepared in
\cite{fri10,fri11} showed the existence of magnetic order at room
temperature. However, and in clear contrast, several studies
reported the existence of superconductivity through measurements
in single nanotubes as well as bundles of them (single- and
multiwall)
\cite{tan01,koc01,lor09,shi12,tak06,mur07,fer06cnt,yan12}.
Apparently, the critical temperature obtained for the CNT depends
on the sample and the experimental method used; it ranges between
$\sim 0.5~$K to $\sim 15~$K.

The origin of the observed superconductivity in CNT is still under
debate. Maxima in the electronic density of states, called van
Hove singularities, have been used as possible origin for the
superconducting-like signals measured after application of a gate
voltage \cite{yan12}, purely electronic mechanism in certain
geometries of ultra-small diameter ($\lesssim 2~$nm) \cite{bel06},
an overwhelming attractive electron-phonon interaction  in
double-wall CNT (DWCNT)~specially when the outer tube is metallic
\cite{nof11}, are some of the theoretical concepts one finds
nowadays in recent studies predicting superconductivity in CNT.

In this experimental work, we studied the electrical transport
properties of bundles of, mostly, DWCNT as a function  of
temperature and magnetic field. The observed negative
magnetoresistance of the CNT-bundle in the whole temperature range
indicates the existence of spin dependent scattering processes up
to room temperature, similar to that found in materials with
defect-induced magnetism (DIM), see e.g. \cite{kha12}. We found
that an extra contribution to the magnetoresistance appears at $T
< 15~$K and at low enough input currents. This fact and the
observed positive magnetoresistance with its hysteretic behavior
suggest the existence of superconductivity in a similar
temperature range  as the one found recently in DWCNT \cite{shi12}
as well as in pyrolytic graphite flakes under an electric field
\cite{bal14}. The temperature dependence of the magnetization, its
hysteresis and other features support also the existence of
magnetic order and to some extent of superconductivity.

\section{Experimental details}
\subsection{Carbon Nanotubes Synthesis}

\begin{figure} 
\centering
\includegraphics[width=1.1\columnwidth]{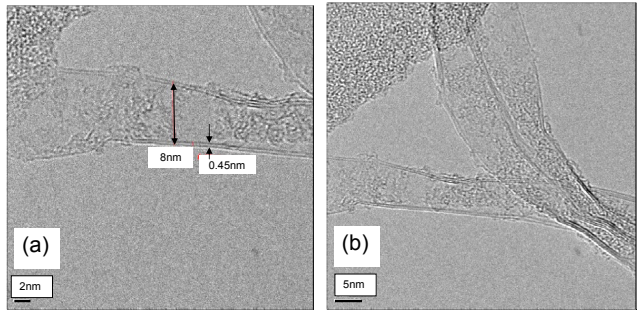}
 \caption{Transmission electron pictures of the measured DWCNT.}
 \label{TEM}
 \end{figure}

The used procedure to synthesize the CNT was the following: The
wafers for the growth substrates were prepared by e-beam
evaporation (base pressure between $7 \times 10^{-7}$ and $2
\times 10^{-6}~$Torr) where 1.2~nm Fe and 10~nm Al$_2$O$_3$ films
were deposited on 4~inches diameter silicon wafers. The gases used
for the carbon nanotubes (CNT) growth were Ar (99.9995\%, Maxima),
C$_2$H$_4$ (99.5\%, Maxima), H$_2$ (99.999\%, Maxima), and a
mixture of 99\% Ar with 1\% oxygen, which we will denote as
Ar/O$_2$ (Maxima). The flows were maintained using electronic mass
flow controllers (MKS, model 1179A). The experiment was performed
on 0.5~cm$^2$ catalyst substrates and with the same gas preheating
and fast heating techniques previously described
\cite{nes09,nes12}. The synthesis were performed using a
fused-silica tube (internal diameter of 22~mm) placed in two
atmospheric-pressure tube furnaces (Lindberg Blue) with controlled
flows of the source gases - Ar, H$_2$, C$_2$H$_4$, and Ar/O$_2$.
The first furnace (set at 770~$^\circ$C) preheated the gases. The
growth substrate was positioned in the second furnace (at
755$^\circ$C) for the annealing and growth steps. The substrate
were inserted into the furnace for a 15 minutes anneal with
100~sccm of Ar and 400~sccm of H$_2$, followed by a 30~minutes
growth cycle with 100~sccm of Ar, 100~sccm of Ar/O$_2$, 400~sccm
of H$_2$ and 200~sccm of C$_2$H$_4$. At the end of CNT growth, the
H$_2$ and C$_2$H$_4$ were turned off and the substrate was
post-annealed in the remaining Ar and Ar/O$_2$ for 2 minutes. Then
the Ar/O$_2$ gas was turned off, and H$_2$ gas was turned on for
an additional minute of anneal. Then the H$_2$ gas was turned off,
and the sample was removed from the heated zone and positioned
above a cooling fan (with Ar still flowing within the tube) until
it was cool enough to be removed for characterization.

\begin{figure} 
\centering
\includegraphics[width=1.0\columnwidth]{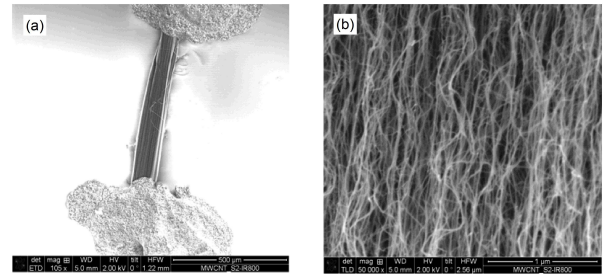}
 \caption{(a) Scanning electron microscope (SEM) picture of one of the
 measured bundles with the voltage electrodes. The scale bar is 500~$\mu$m.
 (b) SEM picture with higher resolution of the CNT bundle. The scale bar is 1~$\mu$m.}
 \label{SEM}
 \end{figure}

\subsection{Characterization methods}
\begin{figure} 
\centering
\includegraphics[width=1.1\columnwidth]{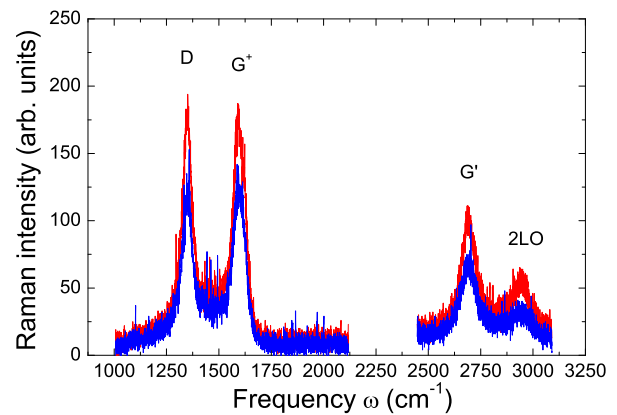}
 \caption{Raman spectra obtained at two different positions of a
 DWCNT bundle (blue and red
 curves).}
 \label{raman}
 \end{figure}

The morphology of the CNT was examined using scanning electron
microscopy (SEM, Quanta FEG 250),
 high-resolution  transmission electron microscopy
(HRTEM, Jeol 2010) and a dual beam microscope NovaLab XT200 from
FEI. Figure~\ref{TEM} shows two TEM pictures where we can
recognize the double-wall and the diameter of the CNT.
Figure~\ref{SEM} shows scanning electron microscope (SEM) pictures
of the measured bundles with the voltage-current electrodes at the
edges (a). In Fig.~\ref{SEM}(b) we can recognize that most of the
CNT are connected. Although we could not measure directly the
transport response of a single interconnection, we speculate that
these may have some contribution on the observed behavior we
describe below. Information about the elemental composition of the
bundles was obtained from energy dispersive X-ray (EDX) analysis.
Any traces of magnetic elements like Fe, Ni, etc., were below the
experimental resolution of 50~ppm. \red{However, from recently
published studies \cite{spe14} we know that EDX analysis is not
really appropriate to find and characterize traces of magnetic
impurities embedded in carbon-based materials. Therefore a further
characterization of the magnetic impurities has been made using
particle induced x-ray emission (PIXE), a method that provides
better resolution and other advantages in comparison to EDX
\cite{spe14}. The PIXE results indicate the existence of Fe
dispersed within the bundle of DWCNT with a concentration
$\lesssim 250~\pm 50~\mu$g Fe per gram of carbon, a concentration
equivalent to $\lesssim 60~\pm 15~$ppm Fe. This concentration is
not relevant for electrical transport measurements, because the
small amount of Fe-based grains are dispersed and likely attached
either at some of the edges and/or at the surface of the DWCNT.
However, if this Fe concentration shows magnetic order, it might
provide a clear contribution to the total magnetization of the
bundles and will be taken into account in the discussion. The
concentration of other magnetic impurities is more than one order
of magnitude below that of Fe and therefore not relevant for the
interpretation of the results.}

Micro-Raman spectrum was obtained at room temperature and ambient
pressure with a Dilor XY 800 spectrometer at 514.53~nm wavelength
and a 2~$\mu$m spot diameter. The incident power was kept at
1.5~mW to avoid any sample damage or laser induced heating
effects. \red{ The Raman spectra obtained at two different
positions of a bundle of CNT are shown in Fig.~\ref{raman}. The
spectra show many peaks; as in graphite and single wall carbon
nanotubes (SWCNT) a D band peak (at $\simeq 1350~$cm$^{-1}$), the
G$^+$ peak (around 1580~cm$^{-1}$, also observed in semiconducting
SWCNT \cite{sou01}), the overtone of the D peak, named G' (at
$\simeq 2690~$cm$^{-1}$) and also observed in semiconducting SWCNT
\cite{sou01} and an overtone longitudinal optic (LO, at $\simeq
2940~$cm$^{-1}$) peak. For our bundles we observe a G$^+$/D ratio
of approximately 1, which is indicative of defective CNT.} This is
consistent with the morphology observed in the SEM/TEM images.

For the electrical resistance measurements we separated carefully
bundles of CNT with length of $\sim 800~\mu$m using non-magnetic
tweezers. Afterwards the bundle was placed on the top of a $5
\times 5~$mm$^2$ silicon (100) substrate covered with a 150-nm
insulating  Si$_3$N$_4$ film. The contacts for the resistance
measurements were done using a commercial silver paste  and gold
wires (25~$\mu$m diameter) in a four-two points configuration (see
Fig~\ref{SEM}(a)). We have done also measurements with the usual
four points to check whether the contact resistance contributed.
The results were similar for both, four-two or four points
configurations. The temperature dependence measurements were done
in a commercial $^4$He-flow cryostat (Oxford Instruments) equipped
with a superconducting magnet with  maximal field $\pm 8~$T.
During the measurements the magnetic field was applied
perpendicular to the bundle main length. High resolution
resistance measurements were done using an AC Bridge (Linear
Research LR-700) in the range of temperature between 2 and 275~K.
The temperature stabilization was better than 4~mK in the whole
temperature range. For the current-voltage ($I-V$) measurements we
used a Keithley DC and AC current source (Keithley 6221) and a
nanovoltmeter (Keithley 2182). For the magnetization measurements
we took a CNT-bundle of mass 1.0~mg and fixed it with grease
\red{in the middle of  a 13.5~cm long previously characterized
pure quartz  rod with negligible background \cite{bar07}}. The
magnetic moment of the sample was measured using a superconducting
quantum interference device magnetometer (SQUID) from Quantum
Design.

\section{Results and discussion}
\subsection{Temperature dependence of the resistance and nonlinear response
at zero applied field}

\begin{figure} 
\centering
\includegraphics[width=1.0\columnwidth]{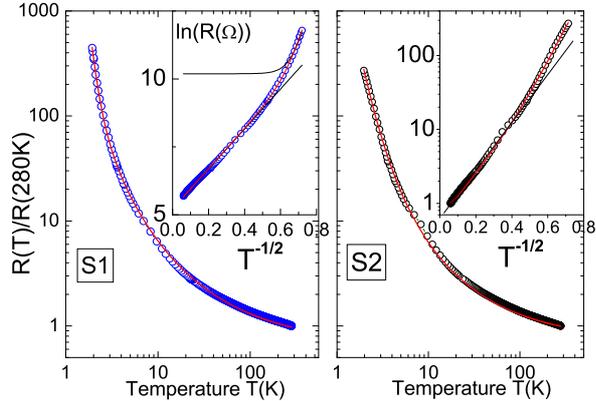}
 \caption{Normalized resistance vs. temperature for two samples, S1 (a) and
 S2 (b)
 at zero applied field and with an input current of $10~\mu$A. The inset in (a) shows the
 natural logarithm of the absolute resistance (in Ohms) vs. the inverse square root of the temperature.
  The inset in (b) shows the normalized
 resistance vs. $T^{-1/2}$ in semilogarithmic scale.
The red lines in the main panels and insets are a fit to
Eq.~(\ref{ad}), which is the addition of the
 two contributions given by Eqs.~(\ref{vrh},\ref{fit}), the two
 black lines in the inset of (a).
 }
 \label{RT}
 \end{figure}

We have investigated in total five samples, which showed similar
behavior. Therefore, we discuss here the results obtained for two
of the samples, S1 and S2. Figure~\ref{RT}  shows the resistance
as a function of temperature for the two samples. The insets show
the same data but vs. $T^{-1/2}$. In the insets we recognize
that the temperature dependence shows two different regions, one
below and the other above $T \sim 10~$K. At temperatures above
$\sim 10~$K, the resistance can be well described by the variable
range hopping theory. In general, the resistance in the variable
range hopping (VRH) regime can be expressed as:
\begin{equation}\label{vrh}
    R_{\rm VRH}(T) = R_0 \exp((T_H/T)^{1/(1+d)})\,,
\end{equation}
where $R_0$ is a free prefactor, $T_H$ is a characteristic
temperature coefficient, and $d$ the dimensionality. In the case
an energy gap is present at the Fermi Level, the VRH resistance
follows $d = 1$ as derived by Efros-Shklovskii \cite{efr75}; for
this special case $T_H = 2.8 e^2/\xi k_B \epsilon$, where
$\epsilon$ is the dielectric constant, $e$ the elementary charge
of the electron, $\xi$ a localization length and $k_B$ the
Boltzmann constant. The insets in Fig.~\ref{RT} indicate that the
resistance of the CNT bundles follows a VRH mechanism (straight
lines in those insets) above $T \sim 10~$K.

The low temperature part of the resistance can be understood using
the fluctuation-induced tunneling (FIT) model, which considers
metallic-like  grains separated by insulating, tunneling barriers
with an effective capacitance \cite{she78,she80}. According to the
FIT model and at small applied electric fields (or currents), the
temperature dependent  resistance across a single, small junction
is given by:
\begin{equation}\label{fit}
    R_{\rm FIT}(T) = R_A \exp(T_1/(T_0 + T))\,,
\end{equation}
where
 \begin{eqnarray}
T_1 &=&  \frac{8\epsilon_0}{e^2k_B}\left ( \frac{A\phi_0^2}{w} \right )\label{t1}\,,\\
T_0 &=& \frac{16 \epsilon_0 \hbar}{\pi \sqrt{(2m)} e^2 k_B} \left
( \frac{A\phi_0^{3/2}}{w^2} \right )\,, \label{t0}
\end{eqnarray}
where $R_A$ is a parameter that depends weakly  on temperature,
$\phi_0$ is the barrier height, $w$ a barrier width and $A$ its
lateral area, which is the effective geometrical area at the
contact between the two metallic regions, $\epsilon_0$ the vacuum
permittivity,  and $m$ the electronic mass. The characteristic
temperature $T_1$ can be interpreted as the energy required for an
electron to cross the barrier $\phi_0$ and $T_0$ a temperature,
well below which, thermal fluctuation effects become negligible.

We found that the simple addition of these two mechanisms, i.e.
assuming the two resistances in series
\begin{equation}\label{ad}
    R(T) = R_{\rm VRH}(T) + R_{\rm FIT}(T)\,,
\end{equation}
fits the experimental data, as shown by the continuous curves
through the experimental points obtained at $I = 10~\mu$A , see
Fig.~\ref{RT}. The parameters that fit the experimental data have
some correlations and therefore should be given within some
confidence range. Similar curves as shown in Fig.~\ref{RT} can be
obtained compensating the effects of the parameters within the
following range: $R_0 \simeq 180 \pm 10~\Omega$, $T_H \simeq 50
\pm 5~$K, $R_A = 1 \ldots 10~\Omega$, $10 \lesssim T_1 \lesssim
30~$K and $0.5 \lesssim T_0 \lesssim 2~$K.

\begin{figure}
\centering
\includegraphics[width=1.0\columnwidth]{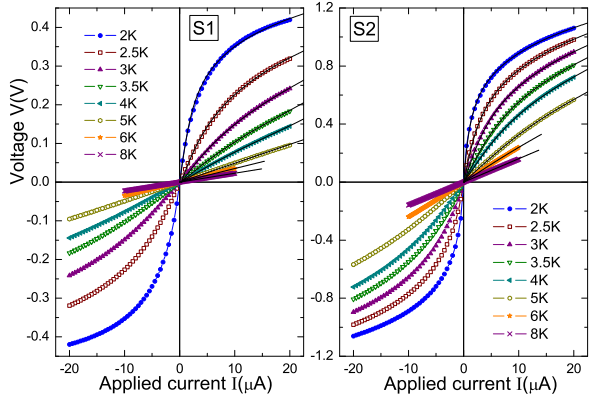}
 \caption{Voltage-current ($V-I$) characteristic curves measured for samples S1 and S2 at
 different constant temperatures and at zero applied field. The continuous lines through
 the experimental data are fits to the function given by Eq.~(\ref{nlt}).}
 \label{IV}
 \end{figure}

Taking into account the SEM pictures, see Fig.~\ref{SEM}, we
expect that the electric current is not being transported by the
same single CNT all along the electrodes but most of them are
interconnected having junctions. In this case the resistance will
be the sum  of the two contributions, one from the 1D transport
through the CNT and the other through the junctions in series. It
is important to note that the FIT model implies that a nonlinear
contribution to the ohmic resistance should exist. This
contribution is given by a current-voltage characteristic of the
type:
\begin{equation}\label{nl}
    I_{\rm FIT}(V,T) = I_0 \exp \left [ - \frac{T_1}{T_0 + T} \left ( 1 -
    \frac{V}{V_0} \right )^2 \right ]\,,
\end{equation}
where $V_0$ is a critical voltage. The measured current-voltage
$(I-V)$ characteristic curves, see Fig.~\ref{IV}, show indeed such
a nonlinear behavior that vanishes at $T \gtrsim 10~$K. Note the
perfect symmetry of the $I-V$ curves for both signs of the applied
current, in contrast to the $I-V$ behavior observed in single CNT
junctions \cite{fuh00}. On the other hand, the nonlinearity of our
$I-V$ curves at $T \lesssim 10~$K is similar to that observed in
ropes of SWCNT in the same temperature range \cite{boc97}.

The experimental $I-V$ curves can be fitted adding the
contribution from the FIT model, Eq.~(\ref{nl}), plus an ohmic
term $I_l = V/R_l$ ($R_l$ is an ohmic resistance), which
corresponds to the electrical paths without barriers:
\begin{equation}\label{nlt}
    I = I_{\rm FIT}(V,T) + (V/R_l(T))\,.
\end{equation}
Figure~\ref{IV} shows the measured $I-V$ curves for samples S1 and
S2 at different constant temperatures. The continuous lines
through the experimental data were calculated following
Eq.~(\ref{nlt}) with parameters similar to those used to fit the
temperature dependence of the resistance, see Fig.~\ref{RT}.
Within the confidence range of the fit parameters $T_1$ and $T_0$
and Eqs.~(\ref{t1},\ref{t0}) we can estimate the width $w$ and
area $A$ ranges of the junctions. Assuming $\phi_0 \sim 0.1~$eV
(see Fig.~\ref{IV}), we obtain $2.4~$nm~$ \lesssim w \lesssim
24~$nm, and $5 \times 10^{-19}~$m$^2$~$\lesssim A \lesssim 1.5
\times 10^{-17}~$m$^2$, values that indicate junctions of several
nm length and width. The values and the spread of the fit
parameters obtained using the FIT model are similar to those found
in the literature \cite{kim01,lai12}.

\subsection{Magnetotransport properties}
\subsubsection{Negative magnetoresistance: Magnetic order contribution }

\begin{figure}
\centering
\includegraphics[width=1.0\columnwidth]{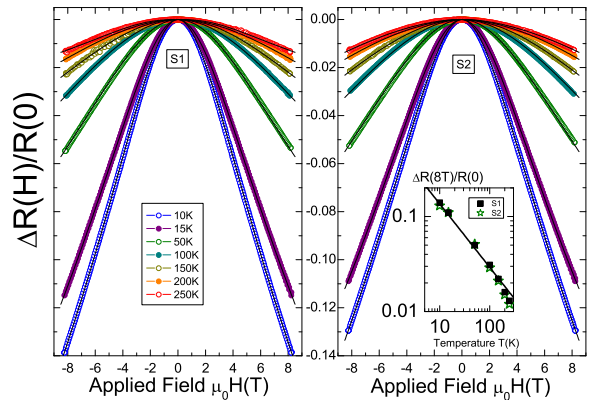}
 \caption{Magnetoresistance for samples S1 and S2 (right panel) as a function of magnetic field
 applied normal to the main axis of the CNT bundles, with a current of $10~\mu$A and at
 different constant temperatures.
 The inset in the right panel  shows the magnetoresistance at a fixed field of 8~T vs.
 temperature in a double logarithmic scale. The straight line follows $0.63 T^{-2/3}$.}
 \label{MR}
 \end{figure}

The measurements of the magnetoresistance (MR) were done at
different applied currents. At temperatures $T \gtrsim 15~$K,
where the VRH behaviour of the resistance overwhelms (see
Fig.~\ref{RT}), the magnetoresistance does not depend on the
applied current for $I \lesssim 20~\mu$A, the maximum current
applied in these studies. In this temperature range the MR is
negative and rather large at low temperatures, see Fig.~\ref{MR}.
Its field dependence can be fitted with the model proposed by
Khosla and Fischer \cite{Kho70} that combines negative and positive
magnetoresistances in semiconductors,  taking into account a
third-order expansion of the $s-d$ exchange Hamiltonian. The
semiempirical formula for the magnetoresistance, defined as
$\Delta R/R(0) = (R(H) - R(0))/R(0)$ as a function of the applied
field $H = B/\mu_0$, is

\begin{equation}
\frac{\Delta R}{R(0)}=
-a^2\ln(1+b^2B^2)+\frac{c^2B^2}{1+d^2B^2}\,, \label{2}
\end{equation}
where $c$ and $d$ are free parameters that depend on the
conductivity and the carrier mobility $\mu$, and
\begin{equation}
a^2 = A_1J\rho_F[S(S+1)+\langle{M^2}\rangle]\,, \label{2a}
\end{equation}
\begin{equation}
b^2 =
\left[1+4S^2\pi^2\left(\frac{2J\rho_F}{g}\right)^4\right]\frac{g^2\mu^2}{(\alpha
kT)^2}\,, \label{2b}
\end{equation}
where  $\alpha$ is a constant. The parameters $a$ and $b$ in
Eq.~(\ref{2}) depend on several factors such as a spin scattering
amplitude $A_1$, the exchange integral $J$, the density of states
at the Fermi energy $N(E_F)$, the spin of the localized magnetic
moments $S$ and the average magnetization square
$\langle{M^2}\rangle$. The negative first term in Eq.~(\ref{2}) is
the term attributed to a spin dependent scattering in third order
($s-d$ in usual $d-$band ferromagnets, $s-p$ in $p-$band
ferromagnets \cite{vol10}) exchange Hamiltonian. The positive
second term in Eq.~(\ref{2}) is a Lorentz-like term that saturates
at high fields and takes into account field induced changes due to
the two conduction bands with different conductivities.

The fits of the experimental data to Eq.~(\ref{2}) for the two
samples are shown in Fig.~\ref{MR}. Although the data can be well
fitted with this model at all measured temperatures, the
correlation and compensation effects between the four free
parameters  is too large to obtain reliable values of the
intrinsic parameters of Eqs.~(\ref{2a}) and (\ref{2b}). Instead,
we show in the inset of Fig.~\ref{MR} the MR at a fixed field of
8~T as a function of temperature for the two samples. The MR
follows very well a $T^{-2/3}$ law at $T < 200~$K. This dependence
suggests that $<M^2(T)> \propto T^{-2/3}$, neglecting the change
in temperature from the parameter $b(T)$ inside the logarithmic
function in Eq.~(\ref{2a}) and the weak contribution of the second
term in Eq.~(\ref{2}). It is reasonable to assume that this has
the same DIM origin as in graphite or nominally non-magnetic
oxides \cite{dim13}. In the case of bulk graphite the dependence
of the ferromagnetic magnetization at fixed fields follows a
nearly linear and negative $T$-dependence, i.e. $M(T) \sim M(0) (1
- aT)$ \cite{barzola2,ram10}. The obtained dependence deviates
clearly from that observed in graphite, probably reflecting the
different dimensionality of the magnetic order in the CNT.

\red{We would like to emphasize that the contribution of the very
small amount of magnetic impurities is unlikely to affect the
magnetoresistance results. The detected $\simeq 60~$ppm
Fe-containing grains would be some at the edges and some at the
surface of the DWCNT. Therefore the main measured voltage due to
the input current should not be influenced by the  impurity
grains. If we compare the case of CNT fully filled with conducting
35~nm diameter Fe reported recently \cite{bar12}, the temperature
dependence of the negative magnetoresistance at a fixed field is
completely different. It remains rather constant between 2~K and
100~K and sharply decreases above this temperature, in clear
contrast to the one measured for our DWCNT bundles, see inset in
Fig.~\ref{MR}. Also, the temperature dependence of the
magnetoresistance of partially Fe-filled CNT \cite{bar12} is
completely different to the one obtained here. Moreover, as we
will show in Section~\ref{magne}, the measured remanent
magnetization of the DWCNT bundle shows a similar temperature
dependence as the one obtained from the magnetoresistance.}

\red{The magnetization data of the DWCNT bundles indicate that,
see Section~\ref{magne}, at temperatures below 300~K a clear
magnetic hysteresis, characteristic of a ferromagnetic state,
which can be well measured. The obtained coercive field amounts to
$H_c \lesssim 0.02~$T at $T \eqslantgtr 2~$K. The question arises
whether a field hysteresis can be measured in the
magnetoresistance. In principle, it should be observed. However,
the magnetoresistance decreases $\propto H^2$ at fields below 1~T.
Therefore, the expected change in the resistance at the measured
coercive fields is $\Delta R(0.02)/R(0) \lesssim 5 \times
10^{-6}$, a change far below our experimental resolution.}

\subsubsection{Nonlinear, positive magnetoresistance}
\label{pmr}

\begin{figure}
\centering
\includegraphics[width=1.0\columnwidth]{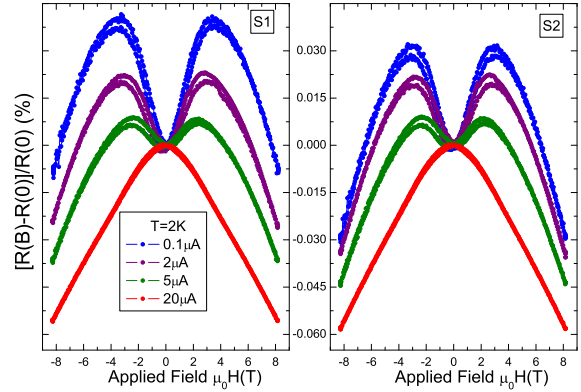}
 \caption{Magnetoresistance of samples S1 and S2 as a function of magnetic field
 applied normal to the main axis of the CNT bundles at 2~K and at different
 currents. Note the development of the positive MR the lower the
 input current as well as the field hysteresis.}
 \label{MR2K}
 \end{figure}

\begin{figure}
\centering
\includegraphics[width=1.0\columnwidth]{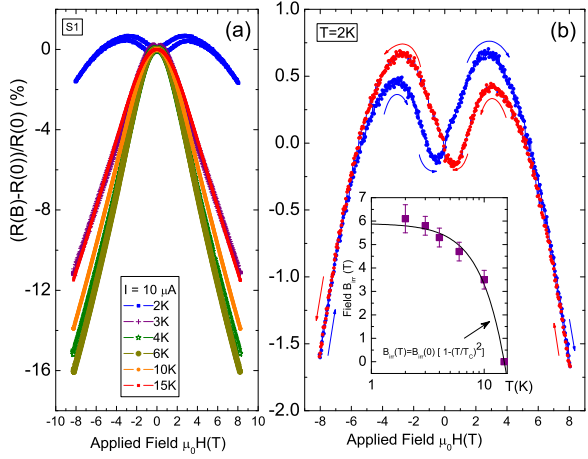}
 \caption{(a) Magnetoresistance of sample S1 as a function of magnetic field
 applied normal to the main axis of the CNT bundles at different temperatures at
 fixed current. Note the development of the positive MR the lower the
 temperature as well as the field hysteresis. (b) The same
 magnetoresistance data as in (a)
but at 2K. The inset shows the temperature dependence of the field
$B_{\rm irr}(T)$ defined as the field within the field hysteresis
loop at which the hysteresis vanishes. The continuous line follows
a simple quadratic $T$-dependence with $B_{\rm irr}(0) = 5.9~$T
and $T_c = 15~$K.}
 \label{Hys}
 \end{figure}

At temperatures $T \lesssim 10~$K,  i.e. in the region where the
nonlinear contribution  that follows the FIT model (see
Fig.~\ref{IV} and Eq.~(\ref{nl})) appears, an extra
 positive MR starts to be measurable as can be
 clearly
 seen in Fig.~\ref{MR2K}. There are a few peculiarities about the
 observed behavior we would like to emphasize:\\

 a) The positive MR is in clear contrast to the
 ferromagneticlike negative MR behavior observed at higher temperatures, compare
 Fig.~\ref{MR} with Fig.~\ref{MR2K}. This {\em positive} and {\em hysteretic} (see Fig.~\ref{Hys}(b))
 MR appears as
 an additive contribution to the
 negative one, similar to the addition of the two contributions to the
 total resistance, see Eq.~(\ref{ad}).

\begin{figure}
\centering
\includegraphics[width=1.0\columnwidth]{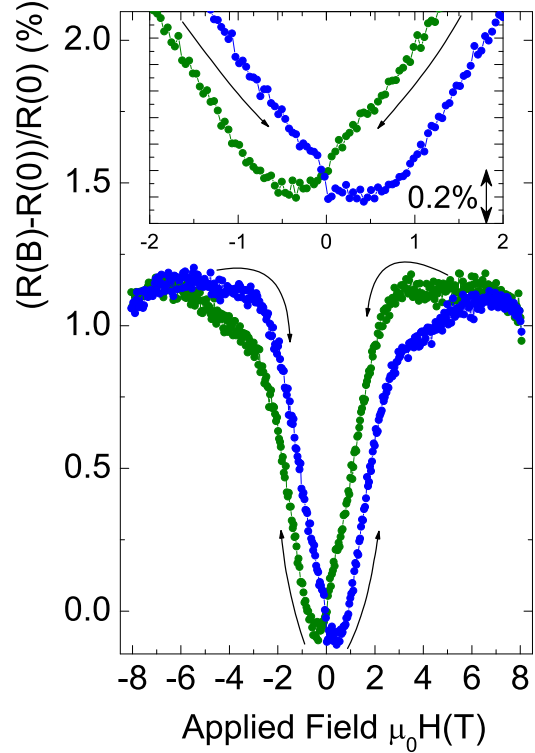}
 \caption{The main panel shows the field hysteresis at $T = 2~$K and input
 current $I = 0.1~\mu$A of sample S1 after subtracting the negative MR contribution following
 a fit to the high field data using Eq.~(\ref{2}), see Fig.~\ref{MR}. The upper inset
 blows up the low field region.}
 \label{Hys2}
 \end{figure}

b) The amplitude of the positive MR and the field hysteresis
decreases with temperature, see Fig.~\ref{Hys}(a).
Figure~\ref{Hys}(b) shows in more detail the hysteresis loop at
2~K and at an input current of $10~\mu$A. As a way to characterize
the temperature dependence of the extra contribution to the MR, we
define the field $B_{\rm irr}(T)$ at which the hysteresis in the
MR vanishes. The temperature dependence of $B_{\rm irr}(T)$ is
shown in the inset of Fig.~\ref{Hys}(b) and follows roughly a
quadratic temperature dependence. This dependence indicates the
vanishing of the hysteresis at a critical temperature $T_c \simeq
15~$K.

c) Figure \ref{Hys2} shows the  magnetic hysteresis of sample S1
at $T = 2~$K after subtraction  of the negative MR following
Eq.~(\ref{2}), as done at high  temperatures. After subtraction,
the MR hysteresis loop shows a behavior compatible to
superconductors, but also to ferromagnets.  The clear addition of
the two contributions
 plus the temperature range where the positive MR and the field
 hysteresis are observed, indicate a relationship of this phenomenon
 with the properties of the junctions.

\begin{figure}
\centering
\includegraphics[width=1.0\columnwidth]{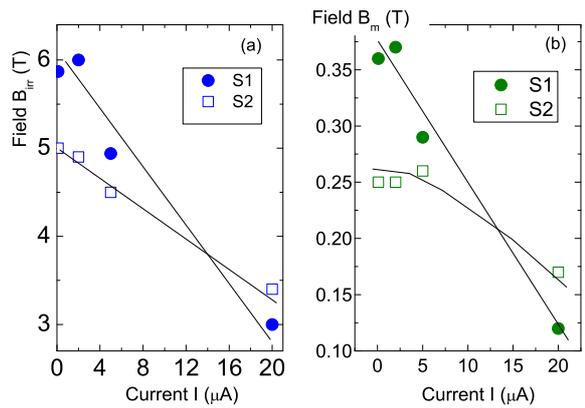}
 \caption{(a) $(\bullet,\square)$: Field at which the field loop gets reversible $B_{\rm irr}$
 vs. applied current,
 for sample S1 and S2, respectively.
 (b) Field at the minimum
of the MR $B_m$, see inset in Fig.~\ref{Hys2},  vs. applied
current for samples S1 and S2 at $T = 2~$K.}
 \label{I-dep}
 \end{figure}

d) The behavior of the positive MR as well as of the field
hysteresis indicate that the input current plays a mayor role in
the development  of this phenomenon. For a better characterization
of its influence we have measured the change of $B_{\rm irr}$ and
the change of the field $B_{m}$ (defined at the minimum of the MR)
with current at a fixed temperature. The results are shown in
Fig.~\ref{I-dep} and indicate a decrease approximately linearly
with current of all these two quantities.

The extraordinary  hysteresis suggests either the existence of
pinning of magnetic entities or a MR hysteresis related to some
spin-valve configurations as observed in single-wall CNT with one
or two magnetically ordered terminals \cite{jen05}. Because we do
not use any magnetically ordered terminals (in our case those are
non-magnetic silver paste) one may think that the magnetically
ordered DWCNT themselves, as the negative MR indicates (see
Fig.~\ref{MR}), act as spin polarized sources. However, it appears
difficult that the DWCNT themselves act simultaneously as
ferromagnetic source and drain with different magnetic
characteristics to produce the necessary hysteresis. In the case
the magnetic entities are magnetic domains existing in the
magnetically ordered DWCNT whose walls can be pinned, it appears
unlikely that the domain walls can produce such large hysteresis
in the MR with a coercive field in the range of $B_m(T=2~$K$)
\simeq 0.35~$T at low input currents, see Fig.~\ref{I-dep}(b).
Note that no measurable hysteresis within experimental resolution
was measured in the negative MR above 15~K.

There are several hints that suggest that the observed positive MR
at low enough currents can be related to a superconducting state,
namely: (1) The clear input current dependence, see
Fig.~\ref{I-dep}, indicates that the positive and hysteretic MR
should be related to the junctions potential barrier that opens
below 15~K. This dependence suggests the existence of a critical
Josephson-like current. The Josephson junctions with their two
superconducting regions should be localized at certain junctions
between DWCNT observed in TEM and SEM measurements.

(2) The value of magnetic field of several Tesla at which the MR
saturates, see for example Fig.~\ref{hys2}, is of the same order
as the upper critical field $H_{c2}$  observed in superconducting
coupled  $4~${\AA} CNT arrays \cite{Wang10}.

(3) The critical temperature  $T_c \simeq 15~$K obtained in our
samples  coincides with several other  $T_c$ reported in
superconducting   $4~${\AA} CNT coupled arrays \cite{Wang10},
superconducting $4~${\AA} CNT/Zeolite composites \cite{lor09}, as
well as DWCNT arrays \cite{shi12}, and last but not least,
 in thin graphite flakes with internal interfaces
 after application of an electric field (with electrical contacts at the top  graphene plane)
 \cite{bal14}.

 (4) Finally, we note that the current necessary to
depress substantially the positive MR and its hysteresis is  $20
\ldots 50~\mu$A, see Figs.~\ref{MR2K} and \ref{I-dep}, which is of
the same order as the current measured in superconducting coupled
$4~${\AA} CNT arrays to reach their normal state \cite{Wang10}.

\begin{figure}
\centering
\includegraphics[width=1.0\columnwidth]{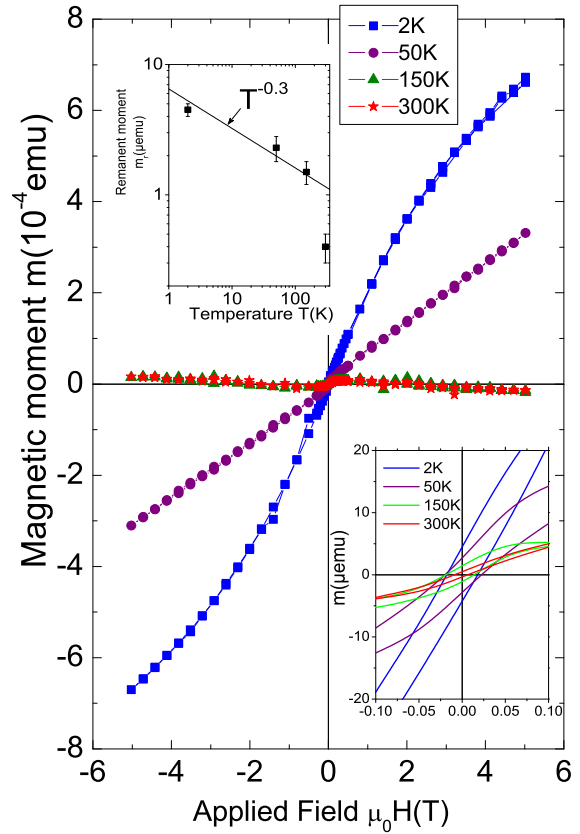}
 \caption{Magnetic field hysteresis for  the sample S1 (mass $\sim 1$~mg)
 measured at different temperatures. \red{Note that due to the used sample holder
 there is no extra magnetic background signal to be subtracted.
 The bottom right inset shows the same data expanding the
 low field region without including the virgin curves for clarity. The top left inset shows the temperature
 dependence of the remanent moment obtained from the field hysteresis curves.}}
 \label{hys1}
 \end{figure}

\subsection{Magnetization results}
\label{magne}
\begin{figure}
\centering
\includegraphics[width=1.0\columnwidth]{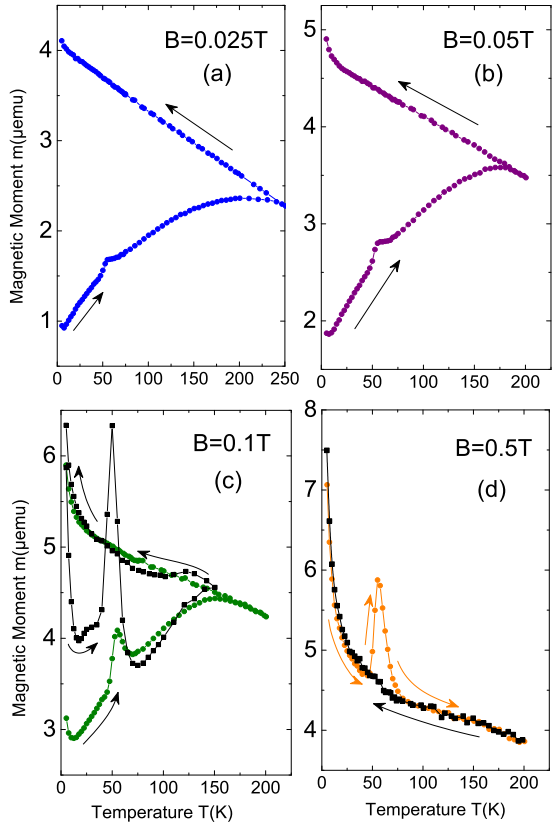}
 \caption{Zero field cooled and field cooled curves (warming up and cooling down respectively),
 of a DWCNT-bundle of sample S2 of mass $\simeq 1~$mg mass,
 at four different fields  applied normal to the main axis of the CNT.
 Two runs taken at $B = 0.1~$T and at different maximum temperatures
 are shown in the corresponding figure as an example of the
 fair reproducibility of the measurements.}
 \label{hys2}
 \end{figure}

\begin{figure}
\centering
\includegraphics[width=1.0\columnwidth]{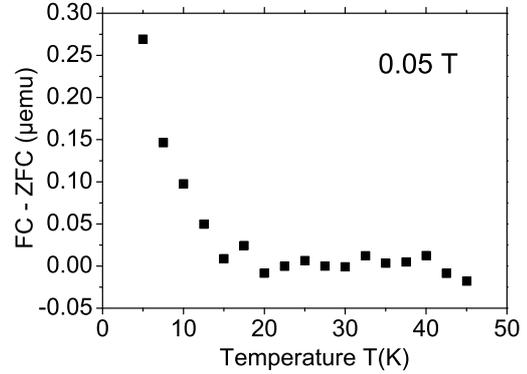}
 \caption{The experimental data points represent
 the calculated difference between the FC and ZFC curves of  sample S1
 obtained at $B = 0.05~$T choosing a maximum temperature of 45~K to avoid the
 influence of the 50K-peak observed in the ZFC curve.
 The background difference from the large ferromagnetic contribution
 has been subtracted using a simple linear in temperature relationship, i.e.
 (FC-ZFC)$_{\rm FM} \simeq 2.3 \times 10^{-8} (45 - T[{\rm K}]) -
 10^{-7}$ [emu].
 }
 \label{hys3}
 \end{figure}

Taking into account that the magnetotransport properties of the
DWCNT  bundles reveal one main contribution, a ferromagnetic-like
in the whole temperature range, we expect that the magnetization
shows a behaviour compatible with it. Figure~\ref{hys1} shows the
field hysteresis of the magnetic moment measured in a bundle of
DWCNT with the field normal to the main axis of the CNT and at
different temperatures. The shown data are as measured, i.e. no
subtraction of any background has been done. \red{Note that the
used sample holder does not show any extra background. In one of
the insets of Fig.~\ref{hys1} we show the same data but in a
expanded field scale. One identifies a coercive field of the order
of 0.02~T at the lowest temperatures.} The magnetization results
indicate a paramagnetic and ferromagnetic contributions with a
ferromagnetic magnetization that saturates at fields of the order
of 0.5~T and remains finite even at room temperature, see inset in
Fig.~\ref{hys1}. \red{Assuming that the detected Fe-concentration
would be ferromagnetic, we expect a maximum magnetic moment at
saturation of the order of $m_{\rm Fe,\rm{sat}}\simeq 50~\mu$emu.
This value is comparable with the saturation magnetic moment of
the DWCNT $m_{\rm{sat}} \sim 60~\mu$emu at 2~K, estimated after
roughly subtracting the large paramagnetic contribution observed
in the DWCNT bundle, see Fig.~\ref{hys1}. On the other side, the
measured paramagnetic contribution clearly overwhelms the largest
contribution expected from the Fe impurities by more than a factor
of ten. Therefore, the  question is, which amount of the
ferromagnetic signal does correspond to the DWCNT and which to Fe?
Without the knowledge of the magnetic properties of the
Fe-containing grains, it is not possible to completely clarify
this question because the behavior of the magnetization depends on
the stoichiometry  (e.g., Fe$_3$O$_4$, Fe$_3$C, etc.)  and size of
the Fe-containing grains. Comparing the transport and
magnetization data related only to the ferromagnetic state, an
interesting similarity can be noted. The remanent magnetic moment
obtained from the field hysteresis at zero field follows a
temperature dependence  $T^{-0.3}$ to $\sim 150~$K, see upper left
inset in Fig.~\ref{hys1}. This is the same temperature dependence
we would obtain for the ferromagnetic magnetization from the
negative magnetoresistance at fixed high fields, see inset in
Fig.~\ref{MR}. This similarity suggests that at least part of the
ferromagnetic signal in the magnetization measurements might come
from the DWCNT themselves.}

It should be clear that the possible signal in the magnetization
coming from the superconducting-like contribution observed at very
low input currents in the MR (see Sec.~\ref{pmr}), is expected to
be small and added to the overwhelming para- and ferromagnetic
contributions. The measured field hysteresis provides us no clear
hints about this extra contribution. Therefore, we decided to
measure the temperature hysteresis, i.e. the zero field cooled
(ZFC) measured by warming and field cooled (FC)  measured
afterwards during cooling at fixed applied fields. The results are
shown in Fig.~\ref{hys2} at four different fields and at maximum
temperatures of 150~K ($B = 0.1~$T), 200~K ($B = 0,05, 0.1,
0.5~$T) and 250~K ($B = 0.025~$T). The main observation we would
like to stress is the large temperature hysteresis that gets
smaller the larger the applied field, in agreement with the
ferromagnetic field hysteresis loop. This hysteresis in
temperature and magnetic field support the existence of magnetic
order in our DWCNT, whatever its origin.

In the ZFC, warming up curves a clear maximum at $T \sim 50~$K is
observed. This maximum, which is a prominent anomaly  that shifts
slightly to higher temperatures with the applied field, is
anomalous because at large enough fields it overwhelms the FC
curve. We do not have an explanation for it but we can rule out
that it is due to the usual oxygen signal from the SQUID cryostat,
since we checked its reproducibility in two SQUIDs and measuring
other samples, which do not show any oxygen peak. The
reproducibility of this peak indicates that, even if this would be
related to oxygen, it should come from the DWCNT bundles. It is
interesting to note that a similar behavior has been reported for
a superconducting amorphous carbon-sulfur (a-CS) powder in
Ref.~\cite{fel122}, see also Figs.~14 and 15 in \cite{fel14}, and
in the same temperature range. In contrast to those reports,
however, the peak in the ZFC curve we observe is always
reproducible in the ZFC curves and does not depend on time and/or
how many times we measured the ZFC curves. Nevertheless the
similarities are remarkable and, as pointed out in
Ref.~\cite{fel14}, the fact that at the maximum the ZFC curve is
above the FC curve, is unique. Even if this maximum would be
related to magnetic impurities, the mechanisms that produce
 such an anomalous behavior remains still unclear.

Due to the large background signal coming from the ferro- and
paramagnetic contributions and the large ZFC peak (and its
irreversibility) at 50~K, we decided to measure the ZFC-FC curves
selecting 45~K as the maximum temperature, applying a relatively
small field of 0.05~T. The idea was to check whether a
superconducting-like difference in the temperature hysteresis is
observed below 15~K. Figure~\ref{hys3} shows the obtained
difference between FC and ZFC curves at 0.05~T and after
subtraction of a linear in temperature background contribution.
The obtained difference starts to increase below $\simeq 15~$K, in
agreement with the results shown in Fig.~\ref{Hys}. We note that
this difference is much smaller than the difference from the
ferromagnetic contribution and the absolute value of the total
magnetic signal. Because the hysteresis in the MR is observed up
to large fields, see Fig.~\ref{Hys2}, one would expect to see it
in the ZFC-FC runs as well. However, the paramagnetic contribution
is several orders of magnitude larger than the expected hysteresis
and, therefore and within SQUID resolution, no clear
superconducting-like hysteresis response could be obtained.

\section{Conclusion and open issues}

Measurements of the magnetotransport properties of bundles of
DWCNT reveal the existence of magnetic order that remains up to
room temperature. At small enough input currents, a positive and
hysteretic magnetoresistance behaviour is observed that vanishes
at $T \simeq 15~$K, compatible with the existence of
superconductivity, in agreement with different reports of CNT as
well as in other carbon-based materials that show a similar
superconducting transition temperature. Although not a
straightforward  proof due to the complexity of
 the observed hysteresis and the possible contribution of Fe
 impurities, we may conclude that the magnetization data appear compatible
to some extent to the interpretation of the magnetotransport data.

Several questions remain unanswered yet. The first is related to
the observed magnetic order. Is it due to defects, like
C-vacancies or to the curvature of the CNT, or hydrogen? Taking
into account the evidence obtained for graphite and theoretical
works, all those possibilities may apply. Future experiments
\red{should try to decrease substantially the amount of Fe
impurities} as well as to measure an isolated DWCNT to study the
change in the magnetotransport properties with annealing and/or
proton bombardment, for example. Second, where is the
superconductivity actually localized and why is it triggered
there? Which kind of magnetic vortices exist that produce such
relatively large hysteresis in the MR? Due to the apparent
relationship of this transition to the development of a tunneling
region, it is tempting to assume that the superconductivity
appears at the junctions of the DWCNT. We note that a junction may
consist of a region of two bilayers graphene, twisted by a certain
angle. In this case, regions with flat bands may appear where
superconductivity develops,  following recently published work on
the extraordinary superconducting properties of graphite
interfaces, see \cite{bal13,esqarx14} and refs. therein. Those
localized regions are expected to be rather small, below 1~$\mu$m
in length, see Figs.~\ref{TEM} and \ref{SEM}. If superconductivity
is localized at these 2D interfaces, can an applied field produce
the so-called Pearl vortices \cite{pea64}, in such a small
junction area? We note that these vortices have the property to
have a giant effective penetration depth $\Lambda \simeq
\lambda^2/d$, where $\lambda$ is the London penetration depth and
$d$ is the thickness of the superconducting layer. As argued in
the original publication \cite{pea64}, it means that they have a
very long range interaction. In this case the bundle of Pearl
vortices may be pinned by the existence of
 pinning centers distributed in a much larger region than the superconducting region
itself, providing such a relatively large MR hysteresis.

The work  in Israel was partially  supported  by the Israel
National Research Center for Electrochemical Propulsion (INREP;
I-CORE Program of the Planning and Budgeting Committee and The
Israel Science Foundation 2797/11). Fruitful discussions with
Prof. Y. Yeshurun from Bar Ilan University are gratefully
acknowledged.

\bibliographystyle{elsarticle-num}

\end{document}